\documentclass[3p,twocolumn,number,sort&compress]{elsarticle-modified}

\pdfoutput=1
\biboptions{}

\usepackage{siunitx}
\usepackage{placeins}
\usepackage{color}
\usepackage{amssymb}
\usepackage{amsmath}
\usepackage[english]{babel}

\DeclareSIUnit\dbm{dBm}
\DeclareSIUnit\db{dB}
\DeclareSIUnit\samples{S}
\newcommand{\textsubscript}[1]{$_{\text{#1}}$}
\newcommand{\seconds}[0]{\second}

\title{Broadband Electrically Detected Magnetic Resonance Using Adiabatic Pulses}

\author{F. M. Hrubesch}
\ead{florian.hrubesch@wsi.tum.de}
\author{G. Braunbeck}
\author{A. Voss}
\author{M. Stutzmann}
\author{M. S. Brandt}
\address{Walter Schottky Institut and Physik-Department, Technische Universit\"at M\"unchen, Am Coulombwall 4, 85748 Garching, Germany}

\begin{document}
\begin{abstract}
We present a broadband microwave setup for electrically detected magnetic resonance (EDMR) based on microwave antennae with the ability to apply arbitrarily shaped pulses for the excitation of electron spin resonance (ESR) and nuclear magnetic resonance (NMR) of spin ensembles. This setup uses non-resonant stripline structures for on-chip microwave delivery and is demonstrated to work in the frequency range from \SI{4}{\mega\hertz} to \SI{18}{\giga\hertz}. $\pi$ pulse times of \SI{50}{\nano\seconds} and \SI{70}{\micro\second} for ESR and NMR transitions, respectively, are achieved with as little as \SI{100}{\milli\watt} of microwave or radiofrequency power. The use of adiabatic pulses fully compensates for the microwave magnetic field inhomogeneity of the stripline antennae, as demonstrated with the help of BIR4 unitary rotation pulses driving the ESR transition of neutral phosphorus donors in silicon and the NMR transitions of ionized phosphorus donors as detected by electron nuclear double resonance (ENDOR). 
\end{abstract}

\maketitle

\section{Introduction}
Electrically detected magnetic resonance (EDMR) \cite{edmr} is a versatile method to characterize point defects and charge transport in inorganic and organic semiconductors \cite{Dersch1983, Schnegg2012, cochrane2014, Malissa2014, Lo2014}. Due to its high sensitivity it allows for the detection of ensembles with as few as 50 spins in spin-pair-based readout processes \cite{mccamey} and of single electron or nuclear spins when single electron transistors are used for readout \cite{Muhonen2014}. The prototype spin pair investigated in the present paper is formed by an unsaturated paramagnetic silicon dangling bond at the Si/SiO\textsubscript{2} interface between Si and its native surface oxide together with a \textsuperscript{31}P donor electron in close vicinity to the interface \cite{Hoehne2010}. If the spin pair is in a parallel spin configuration, recombination of the electron from the donor to the dangling bond state is Pauli forbidden. Using resonant microwave (mw) excitation the Pauli blockade can be lifted by flipping one of the spins involved which leads to increased recombination and a quenching of the (photo-)current through the sample which in turn can be detected electrically.  The characteristic signature of the dangling bond spin is the P\textsubscript{b0} center with an anisotropic g-factor of $g_\parallel = 2.0018$ and $g_\perp = 2.0081$ \cite{Stesmans1998}. The g-factor of the phosphorus donor is $g_{\rm P}=1.9985$ with a hyperfine interaction with the \textsuperscript{31}P nucleus of $A = \SI{117.53}{\mega\hertz}$  and a nuclear g-factor $g_{\rm n} = \SI{2.2601}{}$ \cite{Feher1959}. This leads to an overlap of one of the hyperfine-split phosphorus resonances and the dangling bond resonances at X-band frequencies which hampers experiments performed on these resonances.

The use of broadband microwave delivery structures can mitigate e.g.~the problems caused by this overlap and allows for multi-frequency or frequency-swept spin resonance experiments which would other\-wise require the use of several resonators and multiple cool-down cycles. Broadband microwave striplines have been used successfully for continuous wave (cw) EDMR experiments \cite{Beveren2008, Dreher2011, Klotz2011} and for pulsed experiments on single spin devices \cite{Koppens2006, Dehollain2013}. However, in contrast to resonator-based EDMR experiments these structures exhibit significant inhomogeneities of the microwave magnetic field $B_1$, which are relevant for pulsed EDMR experiments on ensembles. 

Nuclear magnetic resonance (NMR) experiments with surface coils \cite{Bendall1986, Ugurbil1987, Garwood1989, Staewen1990} and recent experiments on superconducting coplanar waveguide resonators \cite{Sigillito2014}, to name but a few, have demonstrated the capability of adiabatic pulses to correct for such $B_1$ inhomogeneities. Here we show that stripline antenna structures in combination with adiabatic pulses can be very successfully used to excite pulsed EDMR as well as pulsed electrically detected electron nuclear double resonance (ENDOR) employing the same antenna for delivery of electron spin resonance (ESR) and NMR frequencies. In order to use adiabatic or optimum control pulses as discussed e.g. in \cite{Nimbalkar2013}, our broadband setup consists of two arbitrary waveform generators in combination with a \SI{10}{\watt} broadband microwave and a \SI{200}{\watt} radiofrequency (rf) amplifier and is able to generate arbitrarily shaped pulses with microwave frequencies between \SI{2}{} and \SI{18}{\giga\hertz} and radiofrequencies from \SI{1}{} to \SI{150}{\mega\hertz} with $\pi$ pulse times as short as \SI{50}{\nano\second} for a microwave power of \SI{100}{\milli\watt} and \SI{100}{\micro\seconds} for a radiofrequency power of \SI{50}{\milli\watt}.

\section{Simulation of the Stripline Short\label{antenna}}
\begin{figure*}[!t]
\centering
\includegraphics{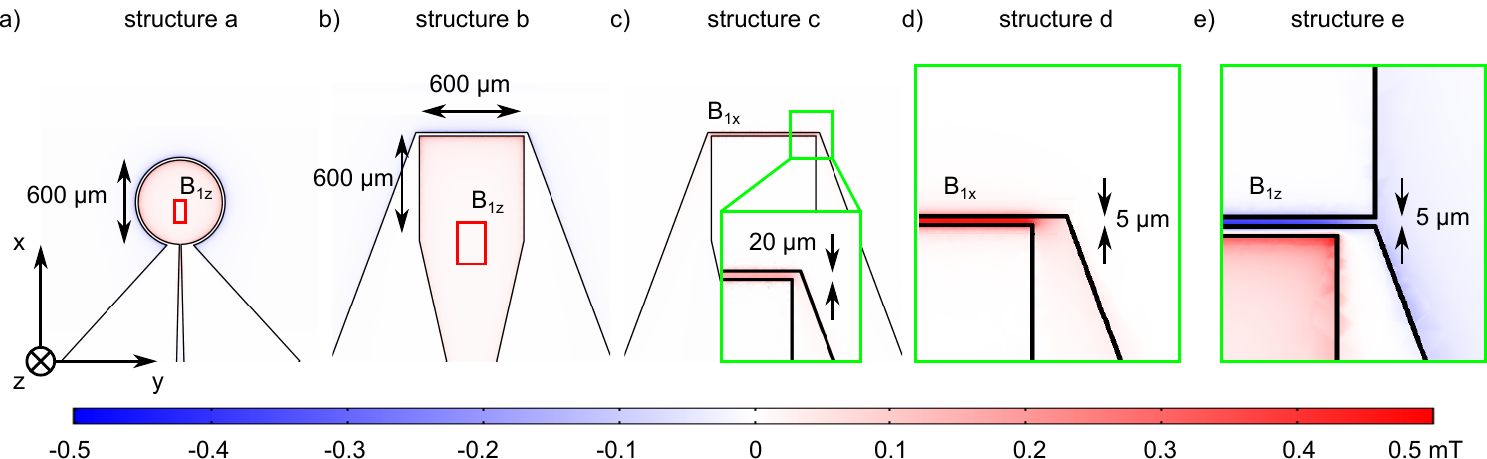}
\caption{Slices through the sample at a frequency of \SI{10}{\giga\hertz} for five simulated structures depicting $B_{1z}$ (panels a, b and e) or $B_{1x}$ (c and d) either \SI{2}{\micro\meter} (panels a, b and c) or \SI{20}{\nano\meter} below the stripline structure (d and e). A voltage of \SI{2}{\volt} corresponding to \SI{40}{\milli\watt} is applied to the coplanar stripline. The red rectangles depict the measurement area of structures a and b. The measurement area for structure c and d is situated beneath the short, whereas the measurement area for structure e is between the short and the additional contact in front of the short.\label{simulation-overview-b-e}}
\end{figure*}

The optimization of stripline structures for single spin experiments has been extensively covered \cite{Dehollain2013}. The requirements stated there also hold for spin ensembles:
The amplitude of the oscillating magnetic field $B_1$ has to be big enough to yield pulse  durations shorter than the spin dephasing time $T_2^*$ of the spins to be studied which is around 60 ns \cite{Lu2011} in our test structures due to the interaction with the \textsuperscript{29}Si nuclear spins present in our samples with natural isotope composition. At the same time the residual electric field has to be kept as small as possible. In addition to these requirements, the homogeneity of the $B_1$ field should be as high as possible to reduce dephasing in the spin ensembles studied. This excludes downscaling of the antennae to the nanometer scale to achieve the highest possible $B_1$ conversion factors, since this reduces the size of the spin ensemble or results in higher $B_1$ inhomogeneities in larger samples. To find the optimum stripline structure for spin ensembles we simulated the electric and magnetic fields at the stripline shorts using the software COMSOL 4.3a. The simulation volume was set to a cube with an edge length of \SI{2}{\centi\meter} and perfectly conducting walls. For the silicon sample a cuboid of the dimensions $\SI{15}{\milli\meter}\times \SI{4}{\milli\meter} \times \SI{0.35}{\milli\meter}$ positioned in the center of the simulation volume was used. On top of the sample the stripline structure was defined by a perfect electric conductor with infinitly small thickness. The electromagnetic wave was excited by a lumped port with a voltage amplitude of \SI{2}{\volt} which corresponds to a microwave power of \SI{40}{\milli\watt} in microwave lines with an impedance of \SI{50}{\ohm}. 

Figure \ref{simulation-overview-b-e} shows slices through the sample at a frequency of \SI{10}{\giga\hertz} for five simulated structures either \SI{2}{\micro\meter} (panels a, b and c) or \SI{20}{\nano\meter} below the stripline structure (d and e) depicting either $B_{1z}$ (a, b and e) or $B_{1x}$ (c and d). In the following, the structures will be called a to e according to their panel labels. The red squares in figure \ref{simulation-overview-b-e}a and b depict the area where a contact structure such as interdigit fingers would be placed for EDMR experiments with a $B_1$ inhomogeneity as obtained from the simulation of less than \SI{\pm5}{\percent}. In figure \ref{simulation-overview-b-e}c the area exactly beneath the short has a similar $B_1$ inhomogeneity of less than \SI{\pm10}{\percent}. In structure d the short width is reduced from \SI{20}{\micro\meter} in structure c to \SI{5}{\micro\meter}. Although fabrication of a contact structure beneath the short is still possible for this reduced width, e.g.~using electron beam lithography, this would increase fabrication complexitiy significantly. Therefore, structure e is intended to be used as microwave delivery and contact structure at the same time. For this, an additional metallization \SI{5}{\micro\meter} in front of  the 5-\si{\micro\meter}-wide short is added. The area beneath the short and the gap between stripline structure and the additional contact in figure \ref{simulation-overview-b-e}e have a higher $B_1$ inhomogeneity of up to \SI{\pm50}{\percent}.

All $B_1$ amplitudes obtained from the simulation quoted below have been divided by a factor of two since for magnetic resonance only one of the two circularly polarized fields contributes to the spin manipulation in the rotating wave approximation. The average $B_1$ amplitude inside the indicated measurement area is \SI{0.02}{\milli\tesla} for structure a and \SI{0.01}{\milli\tesla} for structure b which results in expected microwave power-to-B\textsubscript{1} conversion factors of \SI{0.1}{\milli\tesla\per\sqrt\watt} and \SI{0.05}{\milli\tesla\per\sqrt\watt}, respectively, leading to a power requirement of \SI{13}{\watt} and \SI{50}{\watt} for \SI{50}{\nano\second} $\pi$ pulses, typical for standard commercial pulsed X-band resonators. The average $B_1$ amplitude inside the measurement area underneath the short of structure c is \SI{0.14}{\milli\tesla}, corresponding to a conversion factor of \SI{0.7}{\milli\tesla\per\sqrt\watt} and a theoretical power requirement of \SI{260}{\milli\watt} for a \SI{50}{\nano\second} $\pi$ pulse. By reducing the short width from \SI{20}{\micro\meter} in structures a, b and c to \SI{5}{\micro\meter} in structure d, the average $B_1$ field exactly beneath the short can be increased by a factor of four. With the metal contact in front of the short, a similar average $B_1$ amplitude of \SI{0.58}{\milli\tesla} can also be reached in the gap between the short and the contact. The corresponding conversion factor of \SI{2.9}{\milli\tesla\per\sqrt\watt} should allow for \SI{50}{\nano\second} $\pi$ pulse times with a power of only \SI{15}{\milli\watt}. 

As expected, the simultaneous simulation of the electric field (data not shown) shows that it is smallest at the short where the magnetic field exhibits a maximum and the electric field has a node. Therefore, to reduce the influence of electric fields, the same close proximity of the short and the contact structure already deduced from the optimization of the conversion factor is desirable.  


\section{Broadband EDMR Setup}
\begin{figure*}[!ht]
\centering
\includegraphics[width=\textwidth]{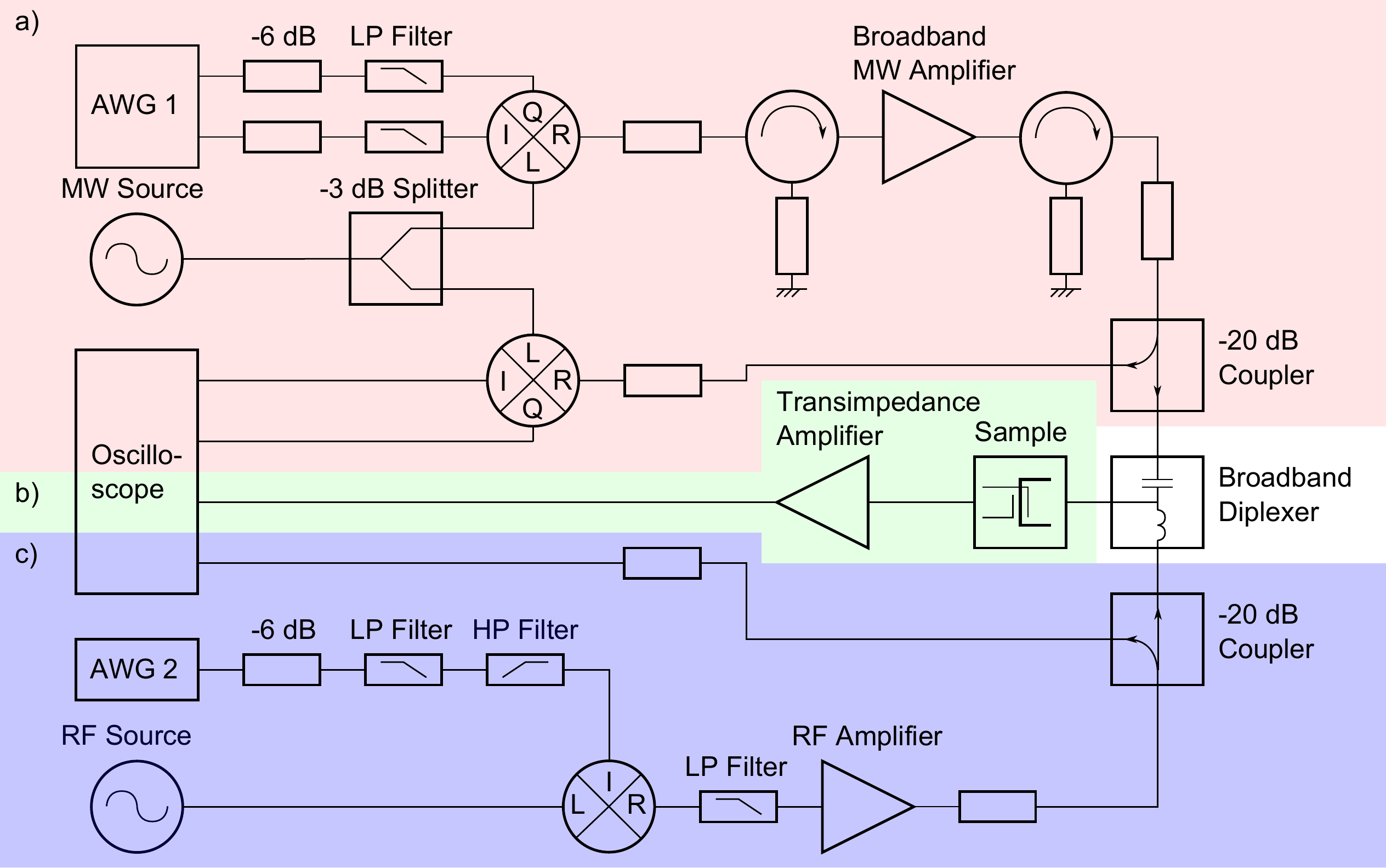}
\caption{Broadband measurement setup composed of the ESR pulse generation (a), the NMR pulse generation (c) and the current measurement (b).\label{setup}}
\end{figure*}

The setup for broadband EDMR using shaped microwave and radiofrequency pulses consists of three parts: the pulse generation for the excitation of ESR transitions which is presented in section \ref{esr-generation}, the pulse generation for NMR transitions (section \ref{nmr-generation}) and the detection circuit, which measures the electric response of the sample to the magnetic resonance excitation (section \ref{detection-circuit}). An overview of the setup is shown in figure \ref{setup}.  

\subsection{Microwave Pulse Generation\label{esr-generation}}
There are two slightly different approaches to ESR spectrometers with pulse shaping capabilites in the literature \cite{Chamm2011, Spindler2012, Doll2013, Kaufmann2013}. We use the approach of \cite{Chamm2011} where pulses are generated at an intermediate frequency of around \SI{100}{\mega\hertz} and an IQ mixer is used as a single sideband upconverter. This method has the advantage that local oszillator (LO) leakthrough can be set to a non-resonant frequency and that the unwanted sideband is suppressed. For optimization, IQ mixers can even be calibrated e.g.~with the method presented in \cite{DeWitt2011} or by minimizing the ESR signal found at the frequencies of the LO or the unwanted sideband.


An arbitrary waveform generator (AWG, Agilent 81180B, sample rate $f_{\rm sample} = \SI{4.6}{\giga\samples\per\second}$) generates the in-phase and quadrature components I and Q of the shaped pulses with a center frequency of $f_{\rm sample}/N$ with $N\in\mathbb{N}$. Those pulses are damped to suppress reflections and low-pass-filtered (MiniCircuits SBLP970) to reduce sampling artifacts before they are upconverted to the desired pulse frequency by the IQ mixer (Marki 0255IMP, Hittite HMC 009 or Marki 0618MMP depending on the frequency range). The cw microwave needed at the LO port of the IQ mixer is provided by a microwave source (Agilent E8257D) and split into two paths by a -\SI{3}{\db} splitter (Marki PD0R618) for coherent up- and downconversion. The pulses at the desired center frequency are damped to adjust their power to the input level of the broadband gallium nitride microwave amplifier (Microsemi AML218P4011, $P_{\rm sat}$=\SI{39}{\dbm} for frequencies from 2 to 18 \si{\giga\hertz}) input. This amplifier can be pulsed in order to reduce the noise during current readout. The isolators (MCLI IS-10, IS-19 and IS-29) at the input and at the output of the broadband power amplifier suppress reflections. The attenuator following the broadband amplifier allows to set the pulse power impinging on the sample. With the -\SI{20}{\db} coupler (Marki C20-0226) a small part of the pulse power is split off for coherent downconversion and pulse analysis. The pulses in the analysis path are damped to the IQ mixer input level and the downconverted pulses are recorded by an oscilloscope (Agilent DSO9254A, sample rate \SI{20}{\giga\samples\per\second}). The bulk of the pulse power reaches a broadband diplexer (SHF DX-45) where the microwave and radiofrequency pulses are combined.

\subsection{Radiofrequency Pulse Generation\label{nmr-generation}}
Although radiofrequency pulses could be synthesized directly, a similiar setup to the microwave pulse generation is used since it enables frequency sweeps without pulse recalculations. In addition it allows to separate the respective AWG2 from the rest of the circuit by pulsing the radiofrequency LO source output which reduces the effect of noise generated by the AWG2 on the detection circuit during readout. The NMR pulses, modulated with a frequency of \SI{150}{\mega\hertz} for mixing with the microwave local oszillator, are synthesized by the AWG2 (Chase DA14000, sample rate \SI{4}{\giga\samples\per\second}), damped to set the pulse power and filtered (MiniCircuits SLP 200+ and SHP 50+) to suppress reflections and remove synthesization artefacts. Afterwards, the pulses are mixed with the pulsed radiofrequency from the rf source (Agilent E4421B) yielding a pulse with the target frequency and a pulse in an unwanted sideband which is filtered by a low pass filter (MiniCircuits SLP 100+). An rf amplifier (ENI 3200L) lifts the pulse power to \SI{200}{\watt} which is subsequently damped to the desired pulse power. A \SI{-20}{\decibel} coupler (MiniCircuits ZFBDC20-61HP+ or ZFBDC20-62HP+) splits the rf into a path for pulse analysis and a path leading to the broadband diplexer. In the analysis path a further attenuator reduces the pulse power to safe levels for the oscilloscope. 

\subsection{Detection Circuit\label{detection-circuit}}
The current through the sample is amplified by a custom-built symmetric transimpedance amplifier (Elektronik Manufaktur Mahlsdorf) with switchable high (\SI{1}{\kilo\hertz}, \SI{5}{\kilo\hertz} or \SI{10}{\kilo\hertz}) and low pass filters (\SI{5}{\micro\second}, \SI{2}{\micro\second} or \SI{<1}{\micro\second}) and an amplification from \SI{1e4}{\volt\per\ampere} to \SI{2e7}{\volt\per\ampere}. The resulting voltage is sampled by the oscilloscope and sent to the measurement computer for further processing.

\section{Measurements}
The following sections are ordered as follows: After a description of the fabrication of the EDMR sample and of the microwave delivery system (section \ref{samplepreparation}), we demonstrate broadband shaped pulse capabilities with structure c in the range from \SI{2.5}{\giga\hertz} to \SI{17.5}{\giga\hertz} (section \ref{broadbandshapedpulse}). Afterwards the conversion factors experimentally realized in structures a, b, c and e are compared to the simulations using Rabi oscillation measurements (section \ref{comparison}). Section \ref{adiabatic-pulses} treats solutions to the high $B_1$ field inhomogeneity using adiabatic pulses. The ENDOR capabilities of the system using shaped pulses for both ESR and NMR are demonstrated in section \ref{endor}.

\subsection{Sample Preparation\label{samplepreparation}}
All measurements were performed on samples from the same silicon-on-insulator wafer which has a \SI{20}{\nano\meter} thick phosphorus-doped Si top layer with natural isotope composition and a doping concentration of \SI{3e16}{\per\cubic\centi\meter} on top of a nominally undoped \SI{2.5}{\micro\meter} thick \textsuperscript{nat}Si buffer layer grown by chemical vapor deposition. The electric measurement structures were defined by photolitho\-graphy and subsequent electron beam evaporation of \SI{20}{\nano\meter} of chromium and \SI{100}{\nano\meter} of gold. The contacts are partially covered by a benzocyclobutene-based photoresist (BCB) to isolate them from the microwave transmission lines which are defined by photolithography, too. They consist of a \SI{50}{\nano\meter} thick chromium layer and a \SI{500}{\nano\meter} thick layer of gold and are connected to the sample holder with four aluminum bonds for each line. The sample holder is made from a Rogers RO3010 microwave printed circuit board with \SI{17}{\micro\meter} thick electrodeposited copper layer and provides a coplanar waveguide (CPW) to coax connection using a Rosenberger 02K243-40ME3 connector and a CPW to coplanar stripline (CPS) transition \cite{Kim2002}.

\subsection{Broadband Capability\label{broadbandshapedpulse}}
\begin{figure}[!t]
\centering
\includegraphics{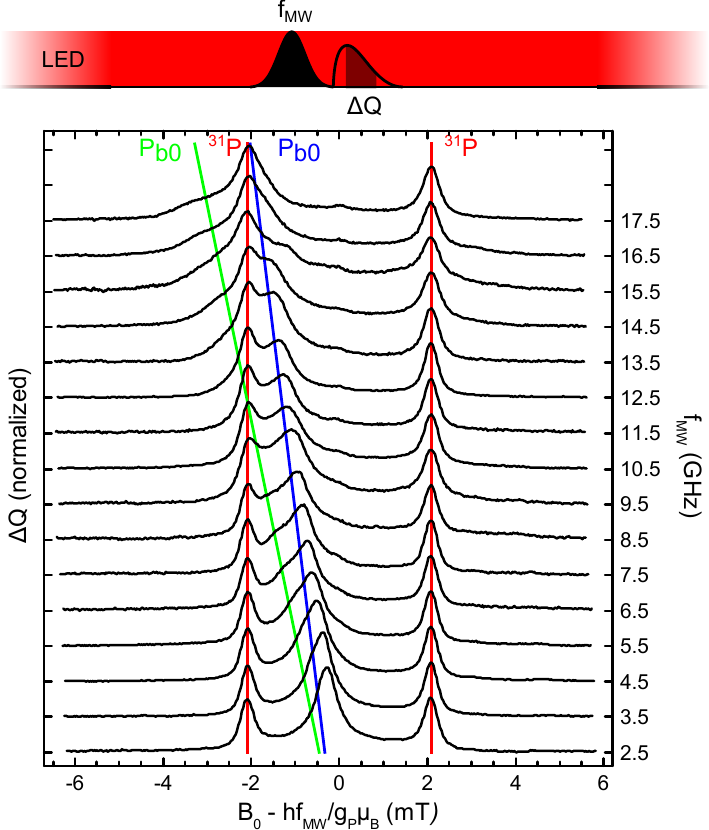}
\vspace{-0.35cm}
\caption{Broadband pulsed EDMR spectra of Si:P for microwave frequencies from \SI{2.5}{\giga\hertz} to \SI{17.5}{\giga\hertz}. The EDMR signal intensity $\Delta Q$ is plotted versus the magnetic field $B_0$ shifted by $\left(hf_{\rm MW}\right)/\left(g_{\rm P}\mu_{\rm B}\right)$ to align the spectra with respect to the donor resonances for different frequencies. The red lines mark the hyperfine-split phosphorus resonances, the blue and green lines follow the two dangling bond resonances observed for the sample oriented with [110] parallel to $B_0$. \label{wasserfall}}
\end{figure}

We first investigate the broadband capabilities of the microwave antennae via Gaussian pulses exhibiting good selectivity \cite{Bauer1984}. The full pulse sequence used for this is shown at the top of figure \ref{wasserfall}. The illumination via a light-emitting diode (LED) is depicted in red and the Gaussian microwave pulse is shown in black. The EDMR response in the form of the charge $\Delta Q$ is obtained through boxcar integration over the current transient after the pulse \cite{Boehme2003, Hoehne2013}. Figure \ref{wasserfall} depicts EDMR spectra for pulse frequencies $f_{\rm MW}$ from \SI{2.5}{\giga\hertz} to \SI{17.5}{\giga\hertz} in steps of \SI{1}{\giga\hertz} measured with a single Gaussian pulse with a standard deviation of \SI{60}{\nano\second} truncated symmetrically to a total pulse length of \SI{400}{\nano\second} which results in a FWHM of \SI{39}{\mega\hertz} in the frequency domain. The ordinate displays the normalized charge $\Delta Q$ while the static magnetic field $B_0$ shifted by $\left(hf_{\rm MW}\right)/\left(g_{\rm P}\mu_{\rm B}\right)$ with Planck's quantum $h$ and  the Bohr magneton $\mu_{\rm B}$ is plotted on the abscissa. This allows a direct comparison of the donor resonances at different microwave frequencies. Figure \ref{wasserfall} clearly demonstrates that pulsed EDMR spectra can be obtained with the antenna structures studied here over the whole range of microwave frequencies between \SI{2}{\giga\hertz} and \SI{18}{\giga\hertz} compatible with the broadband microwave power amplifier. All spectra show a similiar signal-to-noise ratio.

\subsection{Conversion factor of broadband antennae\label{comparison}}
\begin{figure}[!t]
\centering
\includegraphics{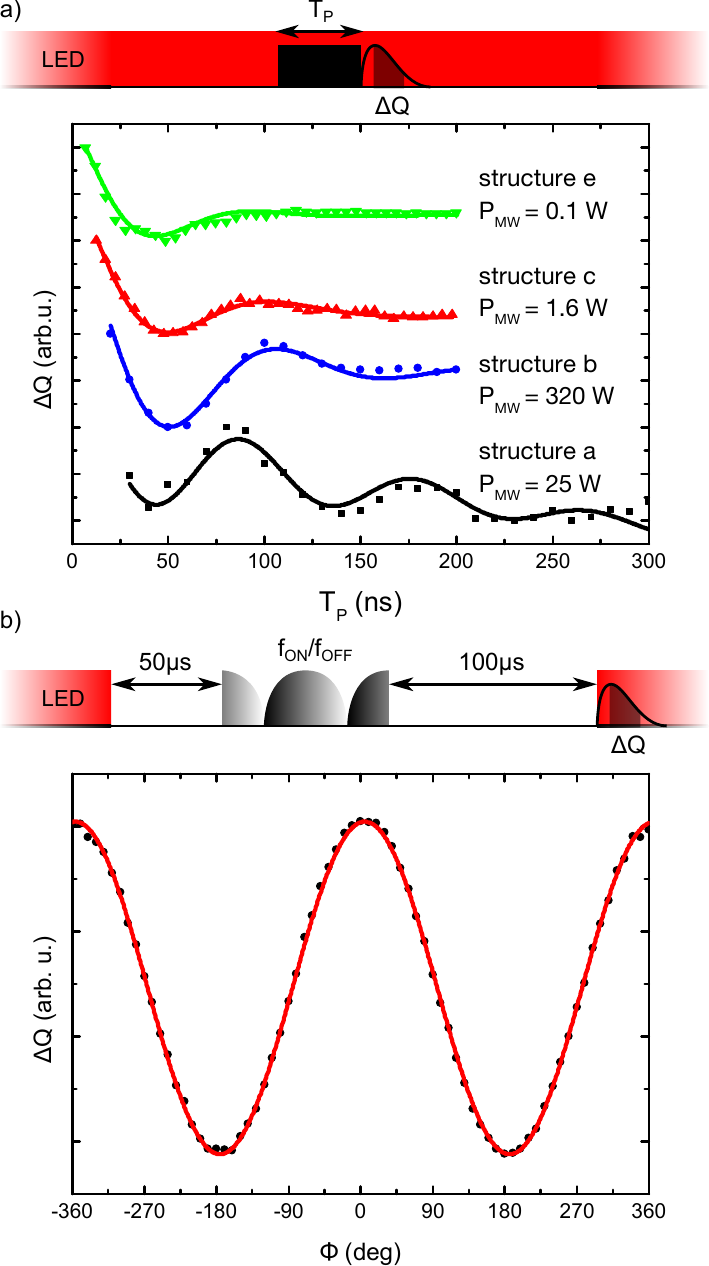}
\caption{(a) Square-pulse-driven Rabi oscillations for structures a, b, c and e measured by EDMR with continous illumination of the sample at a microwave frequency of \SI{9}{\giga\hertz} for structures a and b and \SI{6.5}{\giga\hertz} for structures c and e. The EDMR signal intensity $\Delta Q$ after the mw pulse is plotted versus the pulse length $T_{\rm p}$. (b) Pulse sequence for the study of the BIR4 universal rotation with EDMR. The microwave is applied in the dark, the EDMR is measured during the transient induced by the optical refill pulse. The resulting $\Delta Q$ is plotted versus the rotation angle $\Phi$ of the BIR4 pulse. The red line depicts the expected dependence of the EDMR signal intensity $\Delta Q$ for a BIR4 pulse with varying $\Phi$.\label{rabicomparisonbir4}  }
\end{figure}
In order to experimentally verify the conversion factors of the different structures and compare them to the simulations, Rabi experiments with square microwave pulses have been performed for all structures. The microwave power in each case was set so that the $\pi$ pulse time $T_{\pi}$ is about \SI{50}{\nano\second} corresponding to a $B_1$ of about \SI{0.35}{\milli\tesla}. Since the total pulse times are short compared to the $T_2$ of the system \cite{Huebl2008}, the Rabi experiments have been performed under constant illumination. The corresponding pulse sequence is shown in Figure \ref{rabicomparisonbir4}a together with the results obtained on structures a, b, c and e where the EDMR signal intensity $\Delta Q$ is plotted as a function of the length $T_P$ of the pulse. The data for each structure are fitted with 
\begin{align}
y&=A_{\rm Rabi}\cdot\cos{\left(\frac{2\pi}{T_{2\pi}}\cdot T_{\rm p}\right)}\cdot\exp{\left(-\frac{T_{\rm p}}{\tau_{\rm dep}}\right)}+y_0,\label{rabiformula}
\end{align} 
with the Rabi oscillation period $T_{2\pi}$ and the dephasing time $\tau_{\rm dep}$. This corresponds to a Lorentzian distribution of the $B_1$ fields under the assumption that the g-factor distribution is small in comparison with the $B_1$ inhomogeneity and can be neglected. The half width at half maximum of the $B_1$ distribution can then be calculated as $\Delta B_1 = \hbar /(g\mu_{\rm B}\tau_{\rm dep})$. Used microwave powers $P_{\rm MW}$, Rabi oscillation periods $T_{2\pi}$, the relative dephasing times $\tau_{\rm dep}/{T_{2\pi}}$, relative $B_1$ inhomogeneities $\Delta B_1/B_1$ and the conversion factors $B_1/\sqrt{P_{\rm MW}}$ are listed in table \ref{conversion_factors} for the four antennae.

\begin{table}[!t]
\centering
\begin{tabular}{l|c|c|c|c}
structure&a&b&c&e\\
\hline
$P_{\rm MW}\ [\si{\watt}]$&25&320&1.6&0.1\\
$T_{2\pi}\ [\si{\nano\second}]$&91&112&113&102\\
$\tau_{dep}/{ T_{2\pi}} $&1.44&0.50&0.35&0.30\\
$\Delta B_1/{B_1}$&0.11&0.32&0.46&0.59\\
$B_1/{\sqrt{P_{\rm MW}}}\ \left[\si{\milli\tesla}/{\sqrt{\si{\watt}}}\right]$&0.08&0.02&0.25&1.09
\end{tabular}
\caption{Microwave powers, Rabi oscillation periods, relative dephasing times, $B_1$ field inhomogeneities and conversion factors obtained from the Rabi oscillations in figure \ref{rabicomparisonbir4}a for structures a, b, c and e.}\label{conversion_factors}
\end{table}

Only structures c and e exhibit a large enough conversion factor to generate $\pi$ pulses which are shorter than the $T_2^*$ of our spin ensemble when using a \SI{10}{\watt} broadband amplifier. For the use of the other structures, more powerful amplifiers such as travelling wave tube amplifiers would be needed. The relative increase of conversion factors from structure b to structures c and e are predicted very well by the simulation, although all measured conversion factors are roughly \SI{9}{\dB} lower than the simulated factors. A factor of \SI{3}{\dB} can be attributed to the microwave lines leading to the sample. The remaining difference of \SI{6}{dB} might stem from the assumption of perfect electric conductors for the microwave structure, from the fact that no substrate conductivity was included in the simulation and from reflections at the bond connection between sample holder and sample. In contrast to this, structure a deviates from this scheme with a measured conversion factor that is equal to the simulated conversion factor if the microwave lines are considered. The minimum relative $B_1$ inhomogeneity of \SI{\pm 11}{\percent} calculated from the fit is big compared to the relative $B_0$ inhomogeneity of about \SI{\pm 0.1}{\percent} resulting from the linewidth \cite{Lu2011}, so that the assumption taken in equation \ref{rabiformula} is warranted. A $B_1$ inhomogeneity of \SI{\pm5}{\percent} as calculated for structures a and b should at least result in a $\tau_{\rm dep}/{T_{2\pi}}$ of 3.2. Both structure a and structure b do not quite reach this value with measured values of $\tau_{\rm dep}/{T_{2\pi}} = 1.44$ (structure a) and 0.5 (structure b). The same difference can be found for structure c, where the measured $\tau_{\rm dep}/{T_{2\pi}} = 0.35$ is smaller than the predicted ratio of 1.6. For structure e, where the measured and calculated ratios for $\tau_{\rm dep}/{T_{2\pi}}$ are 0.30 and 0.32, respectively, perfect agreement is found. This behaviour might be caused by the omission of the measurement structures in the simulation of structures a, b and c in contrast to that of structure e, where the microwave delivery itself is used as measurement structure. Since structure e allows to reduce the pulse power to such an extend that the broadband amplifier can even be used in its linear region, all further measurements are performed with structure e.

\subsection{Adiabatic ESR Pulses\label{adiabatic-pulses}}
Using the fit to the Rabi oscillations of structure e we can estimate the negative impact of the dephasing on the modulation depth. At the end of a $\pi$ pulse, the modulation depth $\exp\left(-T_{\rm p}/\tau_{\rm dep}\right)$ is already reduced to \SI{16}{\percent} while the modulation depth for a $2\pi$ pulse is a mere \SI{2.4}{\percent} of the initial EDMR signal intensity. To circumvent the observed effects of the inhomogeneity, an adiabatic half passage pulse with a $\tanh$ amplitude modulation and a $\tan$ frequency modulation was selected
\begin{align}
B_1(t) &= B_{1,\rm max}\cdot\tanh{\left(\zeta\left(1-t/T_{\rm p}\right)\right)}\label{B1-function} \\
\Delta\omega &= \Delta\omega_{\rm max}\cdot\frac{\tan{\left(\kappa t/T_{\rm p}\right)}}{\tan\kappa},\label{f-function}
\end{align}
with constants $\zeta$ and $\kappa$ which determine the adiabaticity of the pulse, the maximum microwave field $B_{1,max}$, the maximum detuning $\Delta\omega_{\rm max}$ and the length of the pulse $T_{\rm p}$. For arbitrary rotation angles this pulse can be concatenated to form the BIR4 composite pulse \cite{Staewen1990}, which is composed of four such adiabatic half passages where the first and the third half passage are applied with a reversed time $t$. The second and the third half passage exhibit an additional phase $\beta$, which sets the rotation angle $\Phi=\left(\beta-\pi\right)\cdot 2$. We use a pulse time $T_{\rm p}$ of \SI{400}{\nano\second} for each adiabatic half passage at an amplitude $B_{1,\rm max}$ of \SI{1.4}{\milli\tesla} (the maximum amplitude which can be generated avoiding non-linearities) and a frequency sweep amplitude $\Delta\omega_{\rm max}$ of $2\pi\cdot 25\ \si{\mega\hertz}$. Due to an increased $T_2$ of the donor electron spin in the dark in our samples, the EDMR experiment is conducted using pulsed light. 
To increase the signal-to-noise ratio a lock-in technique, where sequences with resonant pulses (frequency $f_{\rm on}$) and with offresonant pulses (frequency $f_{\rm off})$ follow each other, is used \cite{Hoehne2012}. With $\kappa = 0.6 $ and $\zeta = 1.45$, which were found by maximizing the adiabaticity 
\begin{align}
\eta(t) &= \frac{\gamma B_{1,\rm eff}}{{\left|d\theta(t)/dt\right|}} \quad\text{with}\\
B_{1,\rm eff}(t) &= \sqrt{B_1^2(t)+\left(\frac{\Delta\omega(t)}{\gamma}\right)^2} \quad\text{and}\\
\theta(t) &= \arctan\left(\frac{\gamma B_1(t)}{\Delta\omega(t)}\right),
\end{align}
where $\gamma$ denotes the gyromagnetic ratio,  $B_{1,\rm eff}$ is the effective $B_1$ amplitude in the rotating frame and $d\theta/dt$ is the rotation speed of the effective $B_1$ in the rotating frame \cite{Garwood2001}, for each half passage, we find a maximum adiabaticity of $\eta = 136$. This extremely high adiabaticity is needed to achieve a satisfactory off-resonance behaviour over the whole linewidth. The full pulse sequence is shown in figure \ref{rabicomparisonbir4}b where the microwave amplitude evolution during the pulse is again depicted by the amplitude and the frequency evolution through the brightness gradient. 

Figure \ref{rabicomparisonbir4}b plots the measured $\Delta Q$ as a function of the rotation angle $\Phi$ of the BIR4 pulse. The measured data is fitted with $\Delta Q=A_{\rm BIR4}\cdot\cos{\Phi}$ which is the expected angular dependence of the recombination on the rotation or rather nutation angle of one of the spin pair constituents. 
The EDMR amplitude when using the BIR4 sequence in figure \ref{rabicomparisonbir4}b is the same as the one determined via a fit to equation \ref{rabiformula} for the square-pulse-driven Rabi oscillations of the same structure e in figure \ref{rabicomparisonbir4}a. Therefore, within the errors of the fits, the BIR4 pulse is able to achieve arbitrary rotations from $-2\pi$ to $2\pi$ for the whole spin ensemble, effectively removing the drawbacks of $B_1$ inhomogeneity from the broadband samples.

\begin{figure*}[!t]
\centering
\includegraphics{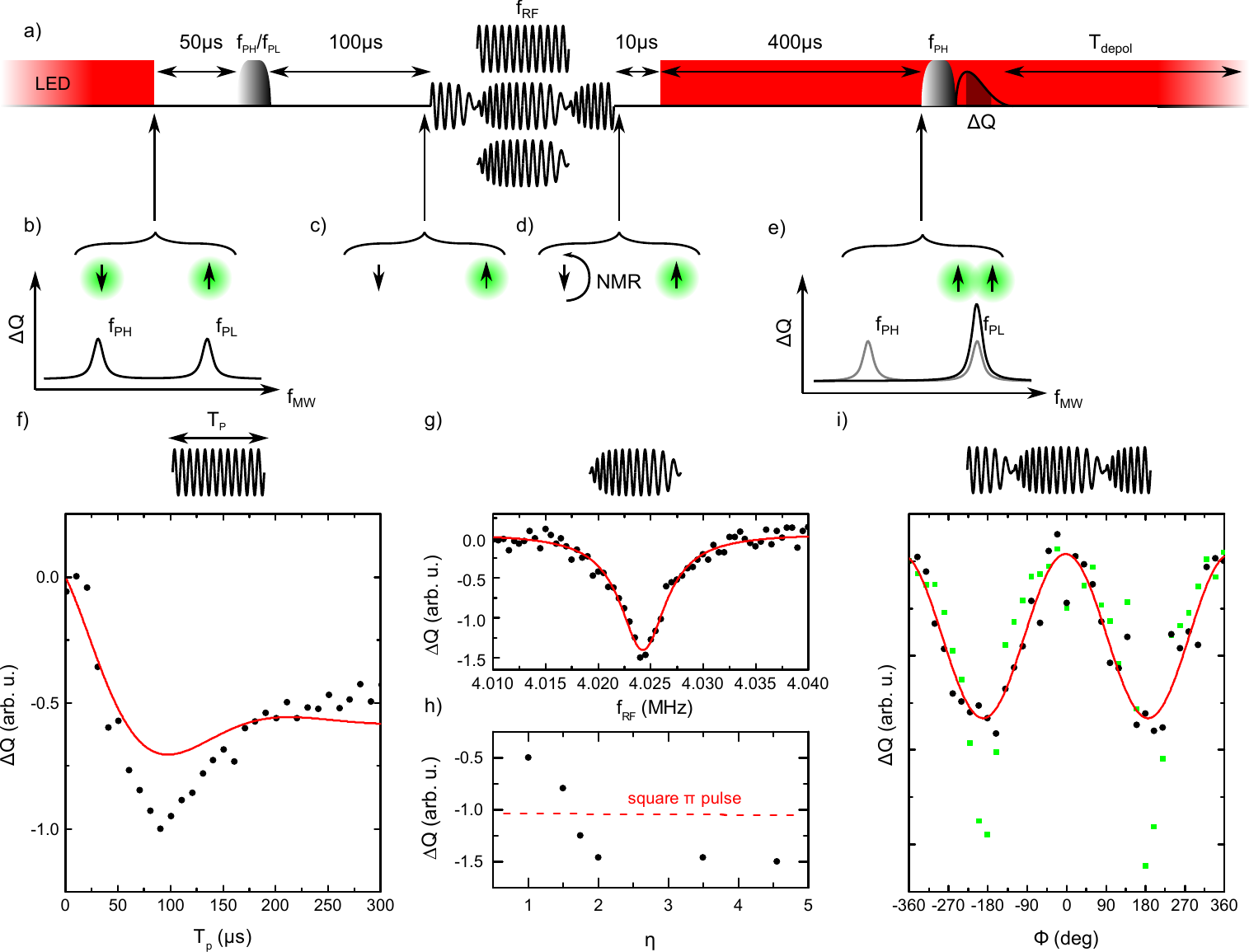} 
\caption{ENDOR experiments using structure e. (a) Modified EDMR Davies ENDOR sequence using adiabatic full passage pulses. Rf pulses are depicted by amplitude- and frequency-modulated sines. (b) Idealized EDMR spectrum of hyperfine-split phosphorus donors as a function of the microwave frequency at the time where the illumination is switched off. (c) State of the nuclear spin ensemble after the ESR full passage donor ionization pulse and a waiting time of \SI{100}{\micro\seconds}. Neutral donors with the donor electron are depicted in green, the ionized donors with the spin of the \textsuperscript{31}P nucleus only. (d) State of the nuclear spin ensemble after application of a ideal inversion pulse on the nucleus via NMR. (e) Idealized EDMR spectrum after all donors have been refilled with charge carriers. (f) Square-rf-pulse-driven nuclear Rabi oscillation for an rf power of \SI{50}{\milli\watt}. The red line is a fit to the data using the dephasing times obtained from the electron Rabi oscillations shown in figure \ref{rabicomparisonbir4}a. (g) ENDOR spectrum using an adiabatic full passage NMR pulse. (h) ENDOR signal amplitude $\Delta Q$ for adiabatic full passage pulses with different adiabaticities $\eta$. The dashed red line indicates the ENDOR signal intensity $\Delta Q$ for a \SI{100}{\micro\second}-long square rf pulse. (i) ENDOR signal $\Delta Q$ as a function of the rotation angle $\Phi$ of a BIR4 pulse for depolarisation times $T_{\rm depol}$ of \SI{2}{\milli\seconds} (green rectangles) and \SI{100}{\milli\seconds} (black dots). The expected dependence of the EDMR signal on the rotation angle $\Phi$ is shown in red. \label{endorrabibir}}
\end{figure*}

\subsection{ENDOR capabilities of the system\label{endor}}
To demonstrate the broadband capability of the delivery system we now turn to ENDOR, using the same antenna and adiabatic pulses for the excitation of the ESR as well as the NMR transitions. We apply a modified Davies ENDOR sequence for EDMR \cite{Dreher2012} based on the difference between the lifetimes of the parallel (about \SI{1}{\milli\seconds}) and the antiparallel (about \SI{10}{\micro\seconds} \cite{Hoehne2013}) state of the phosphorus/dangling bond spin pair to generate a polarization of the ionized nuclear donors. The pulse sequence, which was adapted to use adiabatic full passage pulses for both the ionization and the readout, is shown in figure \ref{endorrabibir}a. Rf pulses are depicted by amplitude- and frequency-modulated sines. For the following discussion we will focus on the high-field hyperfine-split phosphorus electron spin resonance with a nuclear spin quantum number of $m_I=-\frac{1}{2}$ labeled $f_{\rm PH}$ in figure \ref{endorrabibir}b. (The experiment would work the same way if performed on the $m_I=+\frac{1}{2}$ resonance $f_{\rm PL}$.) At the start of the experiment, the light is switched off and during the following \SI{50}{\micro\second} the charge carriers in the bands recombine and the current through the sample settles to zero. All antiparallel donor/dangling bond spin pairs recombine as well. An adiabatic full passage pulse on the $m_I=-\frac{1}{2}$ resonance transforms the remaining donor/dangling bond spin pairs with parallel spin orientation and a $m_I=-\frac{1}{2}$ of the phosphorus nucleus into the short-lived antiparallel state. For the next \SI{100}{\micro\second} these spin pairs also recombine leading to an ensemble of ionized nuclear spins initialized to the $m_I=-\frac{1}{2}$ state  (c.f. figure \ref{endorrabibir}c). Now, NMR on the ionized ensemble can be performed  (c.f. figure \ref{endorrabibir}d) and will lead to a net polarization of all donor nuclear spins depending on the rotation angle of the NMR pulse. Thereafter, the light is switched on for \SI{400}{\micro\second} to allow the recapture of charge carriers at ionized donors \cite{Hoehne2013}. With the full passage ESR readout pulse, the population of the $m_I=-\frac{1}{2}$ state is measured  (c.f. figure \ref{endorrabibir}e). After the readout pulse, the experiment is paused for the time $T_{\rm depol}$ to allow for the light-induced depolarization of the generated nuclear spin polarization which takes about \SI{100}{\milli\seconds} \cite{Hoehne2013a}. 

EDMR relies on the symmetry of a spin pair for its high sensitivity. However, this leads to strong nuclear hyperpolarization \cite{Hoehne2013a}, which makes the analysis of the effects of adiabatic pulses in ENDOR somewhat difficult. If the ionization pulse is switched between the $m_I=-\frac{1}{2}$ and the $m_I=+\frac{1}{2}$ resonances at every other sequence while the readout pulse is kept on the $m_I=-\frac{1}{2}$ resonance, the generated polarization should switch between the $m_I=-\frac{1}{2}$ and $m_I=+\frac{1}{2}$ nuclear spin states, resulting in an alternation of $\Delta Q$ which can be used for lock-in detection. While this lock-in signal will still exhibit some polarization effects, the polarization now mostly depends on the fidelity of the rf pulse and therefore allows for at least qualitative comparisons between different pulses without the need of a light-induced reset of the nuclear polarization.  

All measurements in this section were performed at a $B_0$ field of \SI{234.75}{\milli\tesla}, which results in the microwave frequencies $f_{\rm PH} = \SI{6.5}{\giga\hertz}$ and $f_{\rm PL} = \SI{6.617}{\giga\hertz}$, and with an rf power of \SI{50}{\milli\watt}. Figure \ref{endorrabibir}f depicts nuclear Rabi oscillations driven by square rf pulses with a frequency of \SI{4.0246}{\mega\hertz} after the mw ionization pulse. The measurement was performed using the lock-in technique discussed in the last paragraph with a $T_{\rm depol}$ of \SI{2}{\milli\seconds}. The ENDOR signal intensity $\Delta Q$ is plotted versus the length $T_{\rm p}$ of the rf pulse. Since the polarization effects distort the Rabi oscillation, equation \ref{rabiformula} cannot be used to fit the data in a satisfying way. The red line in figure \ref{endorrabibir}f therefore is a fit of equation \ref{rabiformula} to the data using $\tau_{\rm dep}/{T_{2\pi}}$ from the electron Rabi measurements of structure e. Assuming that the maximum in figure \ref{endorrabibir}f at \SI{100}{\micro\seconds} corresponds to a $\pi$ pulse, we find a $B_1$ field of \SI{0.29}{\milli\tesla} and obtain a conversion factor of \SI{1.3}{\milli\tesla\per\sqrt\watt} close to the conversion factor for the microwave frequencies. The dephasing of the nuclear Rabi oscillation is qualitatively similar to the electron case which is to be expected, because the same antenna and donor/dangling bond ensembles were used. 

In figure \ref{endorrabibir}g the ENDOR signal intensity $\Delta Q$, measured using the lock-in technique with a $T_{\rm depol}$ of \SI{2}{\milli\seconds}, is plotted versus the center frequency $f_{\rm RF}$ of an adiabatic full passage NMR pulse which is concatenated from two adiabatic half passages described in equations \ref{B1-function} and \ref{f-function}. For the adiabatic full passage the parameters $\kappa = 0.62$, $\zeta = 8.6$, found by optimizing the adiabaticity, with a frequency sweep amplitude $\Delta\omega_{\rm max}=2\pi\cdot\SI{8.6}{\kilo\hertz}$ and a pulse length of \SI{100}{\micro\seconds} per adiabatic half passage, were used. With $B_{1,\rm max} = \SI{0.29}{\milli\tesla}$ this results in an adiabaticity factor of $\eta = 2.2$. To test the dependence of the full passage amplitude on the adiabaticity, ENDOR spectra using full passage pulses with different adiabaticities by changing $\kappa$ and $\zeta$ have been recorded. The maximum ENDOR amplitude $\Delta Q$ plotted versus the adiabaticity is shown in figure \ref{endorrabibir}h. The red horizontal line depicts the performance of a $\pi$ square pulse from figure \ref{endorrabibir}f. As can be seen from figure \ref{endorrabibir}h, for adiabaticities higher than 2, $\Delta Q$ saturates. Compared to a rectangular $\pi$ pulse, the  signal intensity is increased by a factor of 1.5.  

As was the case for the electrons, we also concatenated four adiabatic half passages to form the BIR4 pulse to realize universal rotation pulses for the nuclear spin. The length of each half passage was \SI{400}{\micro\second} and we used the parameters $\kappa = 1.09$ and $\zeta = 8.61$ which were optimized for a pulse time of \SI{400}{\micro\seconds} and a sweep width of $\Delta\omega = 2\pi\cdot\SI{8.4}{\kilo\hertz}$ at a $B_{1,\rm max} = \SI{0.29}{\milli\tesla}$ resulting in an adiabaticity factor $\eta=13$. Compared to the BIR4 pulse on the ESR resonances, the required power is much smaller due to the small NMR linewidth of \SI{290}{\hertz} \cite{Dreher2012}. Similar to the nuclear Rabi measurements, the lock-in-detected BIR4 pulse shows polarization if short relaxation times $T_{\rm depol} = \SI{2}{\milli\second}$ are used (c.f. green squares in figure \ref{endorrabibir}i). By increasing the time $T_{\rm depol}$ to \SI{100}{\milli\seconds}, every new experiment starts with a much reduced hyperpolarization of the \textsuperscript{31}P nuclei, and the expected response to the  BIR4 pulse is recovered (black dots).

\section{Conclusion}
We have demonstrated the capabilities of broadband EDMR using shaped pulses for a frequency range from \SI{4}{\mega\hertz} to \SI{18}{\giga\hertz}. The high conversion factor of $\approx$\SI{1.1}{\milli\tesla\per\sqrt\watt} allows to generate \SI{50}{\nano\second} $\pi$ pulses for ESR with a microwave power of $\approx$\SI{100}{\milli\watt} which is well below the amplifier \SI{1}{\dB} compression point of the broadband amplifier. We furthermore demonstrated the delivery of shaped pulses to the broadband antennae as well as the realization of full adiabatic inversion pulses and BIR4 universal rotation pulses for both ESR and NMR, ultimately achieving electrically detected ENDOR with adiabatic pulses only. This opens a wealth of applications for broadband EDMR, ranging from the implementation of optimum control pulses to nuclear polarization shelving in heavier donors in silicon and to the study of spin-dependent charge transport in solar cells made of amorphous or microcrystalline silicon or of organic semiconductors. 

\section*{Acknowledgment}
The authors gratefully acknowledge discussions with D.P.~Franke, S.J.~Glaser, W.~Kallies and R.~Sch\"onman. This work was supported by DFG through SPP 1601 (BR 1585/8) and SFB 631. 

\bibliographystyle{elsarticle-num}
\bibliography{paper}

\end{document}